\documentclass[aps,pra,superscriptaddress,showkeys]{revtex4}

\usepackage{epsf}
\usepackage{graphicx}

\providecommand{\openone}{\leavevmode\hbox{\small1\kern-3.8pt\normalsize1}}

\begin{document}


\title{Memory effects in a Markov chain dephasing channel}

\author{Antonio D'Arrigo}
\email{Antonio.Darrigo@dmfci.unict.it}
\affiliation{MATIS CNR-INFM, Catania 
         \& Dipartimento di Metodologie Fisiche e 
         Chimiche per l'Ingegneria, Universit\`a degli Studi di Catania,  
         Viale Andrea Doria 6, 95125 Catania, Italy}

\author{Elena De Leo}
\affiliation{MATIS CNR-INFM, Catania 
         \& Dipartimento di Metodologie Fisiche e 
         Chimiche per l'Ingegneria, Universit\`a degli Studi di Catania,  
         Viale Andrea Doria 6, 95125 Catania, Italy}

\author{Giuliano Benenti}
\affiliation{CNISM, CNR-INFM \& Center for Nonlinear and Complex systems,
Universit\`{a} degli Studi dell'Insubria, via Valleggio 11,
I-22100 Como, Italy}
\affiliation{Istituto Nazionale di Fisica Nucleare, Sezione
di Milano, via Celoria 16, I-20133 Milano, Italy}

\author{Giuseppe Falci}
\affiliation{MATIS CNR-INFM, Catania 
         \& Dipartimento di Metodologie Fisiche e 
         Chimiche per l'Ingegneria, Universit\`a degli Studi di Catania,  
         Viale Andrea Doria 6, 95125 Catania, Italy}

\begin{abstract}
We study a dephasing channel with memory, modelled by a Markov chain. 
We show that even weak memory effects have a detrimental impact
on the performance of quantum error correcting schemes designed for 
uncorrelated errors. We also discuss an alternative scheme that 
takes advantage of memory effects to protect quantum information.
\end{abstract}

\keywords{Quantum channels; non-Markovian noise; 
quantum error correction}

\maketitle

\section{Introduction}	
 Quantum mechanics offers new and attractive perspectives for information 
 processing and transmission. A large scale quantum computer, if constructed,
 would advance computing power much beyond the capabilities of classical 
 computation,
 while quantum cryptography permits a provable secure data 
 exchange\cite{NielsenChuang,BenentiCasatiStrini}.
 However, due to the unavoidable coupling of any quantum system to
 its environment, decoherence effects appear. This introduces
 noise, thus disturbing the programmed quantum coherent evolution. 
 
 The decoherence problem is conveniently formulated 
 in terms of quantum operations\cite{NielsenChuang,BenentiCasatiStrini}. 
 Given the initial state $\rho$ of a quantum system $\cal Q$ and 
 an overall unitary evolution $U$ of system plus environment, 
 the final system's state $\rho^\prime$ is obtained
 after tracing over the environment
 degrees of freedom: $\rho^\prime=
 {\cal E}(\rho)={\rm Tr}_E[U(\rho\otimes w_0)U^\dagger]$,
 where $w_0$ is the initial state of the environment (we assume 
 that initially the system and the environment are not entangled) 
 and map ${\cal E}$ is known as a quantum operation or a superoperator. 
 It is interesting to consider $\cal E$ as a 
 {\it quantum channel}. This approach encompasses both noisy propagation
 in time and in space. In the first case, map ${\cal E}$ describes the 
 evolution from time $t_i$ to time $t_f$ of some piece of quantum hardware,
 $\rho$ an $\rho'$ being the system's states at $t_i$ and $t_f$,
 respectively.  In the latter, the quantum system $\cal Q$
 plays the role of {\it information carrier} in a two-party communication 
 scenario: $\rho$ is the quantum state at the entrance of the communication
 channel and $\rho^\prime$ the output state, corrupted by noise effects
 described by the quantum operation $\cal E$.
 The {\it quantum capacity} of the channel ${\cal E}$ is the 
 maximum amount of quantum information that 
 can be {\it reliably} transmitted per channel use,
 in the {\it asymptotic limit} of a number $N\to\infty$ of channel 
 uses\cite{BarnumNielsenSchumacher98}.

 Usually quantum channels are assumed to be memoryless, that is, 
 the effect of the environment on each information carrier 
 is always the same and described by map ${\cal E}$. 
 In other words, there is no memory in the interaction between carriers
 and environment: the quantum operation for $N$ channel uses is given by 
 ${\cal E}_N={\cal E}^{\otimes N}$.
 However, in several physically relevant situations this is not a 
 realistic assumption.
 Memory effects appear when the characteristic 
 time scales for the environment dynamics are longer than the time between 
 consecutive channel uses.
 For instance, solid state implementations, which are the most promising 
 for their scalability and integrability, suffer from low frequency 
 noise\cite{kn:solid_state_environment_noise}. In optical fibers,
 memory effects may appear due to    
 slow birefringence fluctuations\cite{Banaszek04}. 
 This introduces correlation among channel uses, {\it i.e.}, 
 the effect of the
 environment on one carrier depends on the past interactions between 
 the environment itself and the other carriers. This kind of channels 
 are known as {\it memory channels}\cite{BowenMancini04,MemoryChannel,KretschmannWerner05}.
 
 A very interesting question, raised for the first time in 
 Ref.~\cite{MacchiavelloPalma02}, is whether memory can {\it enhance}
 the transmission capacity of a quantum channel. This issue is
 relevant also for the performance 
 of Quantum Error-Correcting Codes (QECCs). 
 Since quantum capacity is the maximum rate of reliable 
 quantum information transmission, it puts 
 an upper bound to the asymptotic rate achievable by any QECC. 
 On the other hand, realistic 
 QECCs necessarily work on a finite number of channel uses. 
 Moreover, present day experimental implementations\cite{Cory98,Leung99} 
 are bases on very few channel uses.
 Previous studies have investigated the impact of correlations on the 
 performance of QECCs\cite{QECCcorrelations}. Depending on the chosen model, 
 correlations may have positive or negative impact on QECCs.
 
 In this paper, we consider a Markov chain dephasing channel. 
 In the system-environment Hamiltonian (written in the interaction 
 picture with respect to the system), the $l$-th carrier (qubit) 
 interacts with the environment by means o 
 a single Pauli operator, $\sigma^{(l)}_z$.
 Dephasing channels are thus characterized by the property that, when 
 $N$ qubits are sent through the channel, the states of a preferential 
 orthonormal basis $\{|j\rangle \equiv |j_1,....,j_N\rangle, 
 \,j_1,...,j_N=0,1\}$, with $\{|j_l\rangle\}$ eigenvectors of
 $\sigma^{(l)}_z$, 
 are transmitted without errors. 
 Of course {\it superpositions} of basis states may decohere,
 thus corrupting the transmission of quantum information.
 Dephasing channels model all systems in which
 relaxation times are much longer with respect to dephasing 
 times\cite{kn:solid_state_environment_noise,Leung99}.
 We describe memory effects by means of an irreducible and aperiodic Markov chain. 
 This allows us to find a simple analytical expression 
 for the quantum capacity\cite{Hamada,PlenioVirmani07_1,DarrigoBenentiFalci07_1}, 
 which turns out to be enhanced by memory.
 Furthermore, this model is very convenient to
 illustrate the impact of memory on the 
 performance of the standard three-qubit QECC as well as 
 to compare this three-qubit code with an alternative two-qubit 
 coding/decoding scheme that takes advantage of 
 memory\cite{DarrigoBenentiFalci07_2}.

\section{The model}
 A single use dephasing channel has a very simple 
 representation\cite{NielsenChuang,BenentiCasatiStrini}:
 \begin{equation}
     \rho^\prime={\cal E}\big(\rho\big)=
     \sum_{m\in \{0,z\}} p_m B_m\, \rho^{\cal Q} \,B_m^\dag, 
     \qquad  B_m=\,\sigma_m,
  \label{DephCh_Kraus}
 \end{equation} 
 where $\sigma_0=\openone$.
 Channel (\ref{DephCh_Kraus}) has an intuitive meaning: it 
 leaves the qubit unchanged with probability $p_0$, while it introduces 
 a {\it phase-flip} error with probability $p_z=1-p_0$.
 The generalization to $N$ uses is straightforward:
 \begin{equation}
  \rho_N^\prime= \mathcal{E}_N(\rho_N)=\sum_{i_1,...,i_N} p_{i_1...i_N}\,
  B_{i_1...i_N} \rho_N B^\dagger_{i_1...i_N},
  \quad \quad i_k=0,z,
  \label{dephmemory}
 \end{equation}
 where $\rho_N$ ($\rho_N^\prime$) 
 describes a $N$-qubit input (output) state, and the operators 
 $B_{i_1...i_N}$ are defined in terms of the Pauli operators 
 $\sigma_0= \openone$ and $\sigma_z$: 
 \begin{equation}
  B_{i_1...i_N}\equiv\sigma_{i_1}^{(1)}\otimes
  \cdots\otimes\sigma_{i_N}^{(N)},
  \label{Krausnuses}
 \end{equation}
 with $\sum_{\{i_k\}} p_{i_1...i_N}=1$ and 
 $\sigma_{i_k}^{(k)}$ acting on the $k$-th qubit.
 The quantity $p_{i_1...i_N}$ can be interpreted as the probability 
 that the ordered sequence 
 $\sigma_{i_1}^{(1)},...,\sigma_{i_N}^{(N)}$ 
 of Pauli operators is applied to the $N$ qubits crossing the channel.
 We suppose that probability $p_{i_1...i_N}$ is stationary:
 $p_{i_q}=\sum_{\{i_k, k\neq q\}} p_{i_1...i_N}, \quad 
 \{p_{i_q}\}=\{p_0,1-p_0\}$ for all $q=1,\dots,N$. 
 Memory is introduced by assuming that the joint probability
 $p_{i_1...i_N}$ cannot be factorized:
 $p_{i_1...i_N} \neq p_{i_1}p_{i_2} \ldots p_{i_N}$.
 To describe the joint probabilities in (\ref{dephmemory}) we choose
 a Markov chain:
 \begin{equation}
  p_{i_1,...,i_N}=p_{i_1}p_{i_2|i_1}\cdots p_{i_N|i_N-1},
  \quad \textrm{where} \quad
  p_{i_k|i_{k-1}}=(1-\mu)\,p_{i_k}+\mu\,\delta_{i_k,i_{k-1}}.
  \label{eq:markov-propagator}
 \end{equation}
 Here $\mu\in[0,1]$ measures the partial memory of the channel: it is the probability
 that the same operator (either $\openone$ or $\sigma_z$) 
 is applied for two consecutive uses of the channel, 
 whereas $1-\mu$ is the probability that the two operators are uncorrelated.
 The limiting cases $\mu=0$ and $\mu=1$ correspond to memoryless channels 
 and channels with perfect memory, respectively. In this noise model $\mu$ 
 might depend on the time interval between two consecutive channel uses. 
 If the two qubits are sent at a time interval $\tau \ll \tau_c$,
 where $\tau_c$ denotes the characteristic memory time scale for the environment,
 then the same operator is applied to both qubits ($\mu=1$), 
 while the opposite limit corresponds to the memoryless case ($\mu=0$).
 For this model correlations among different uses decay exponentially,
 and it can be proved\cite{BowenMancini04,DarrigoBenentiFalci07_1,PlenioVirmani07_2} 
 that the channel is {\it forgetful}~\cite{KretschmannWerner05,PlenioVirmani07_2}.
 This property allows us to use the quantum noisy channel coding 
 theorem\cite{BarnumNielsenSchumacher98,CodingTheorem} 
 to compute the quantum capacity of this 
 channel\cite{PlenioVirmani07_1,DarrigoBenentiFalci07_1}: 
 \begin{equation}
  Q=1-p_0H(q_0)-p_zH(q_z),
  \label{capacitymarkov}
 \end{equation}
 where $q_{0,z} \equiv (1-\mu)p_{0,z}+\mu$ are the conditional 
 probabilities that the channel acts on two subsequent qubits via
 the same Pauli operator, and $H(q_0)$, $H(q_z)$ are binary 
 Shannon entropies, defined by $H(q)=-q\log_2 q -(1-q)\log_2(1-q)$. 
 It is interesting to point out that $Q$ increases for increasing 
 degree of memory of the channel. 

\section{Three qubit code: performance in presence of memory}
 Quantum noisy channel coding theorem guarantees, that asymptotically in the 
 number $N$ of channel uses, it exists a QECC code which achieves the quantum capacity 
 (\ref{capacitymarkov}).
 An interesting question is whether memory can be advantageous also 
 for realistic QECCs acting on a small number of channel uses.
 Hereafter we compare the behaviour of two simple coding/decoding schemes 
 for the Markovian dephasing channel (\ref{dephmemory}), the 
 {\it three-qubit code}\cite{NielsenChuang,BenentiCasatiStrini},
 designed for memoryless channels and from now on called code 1 or {c1}, 
 and a {\it two-qubit code}\cite{DarrigoBenentiFalci07_2}, 
 from now on called code 2 or {c2}, that exploits memory effects.
 
 A proper way to measure reliability of quantum information
 transmission is the {\it entanglement fidelity}\cite{BarnumNielsenSchumacher98}.
 To define this quantity we look at the system ${\cal Q}$
 as a part of a larger quantum system ${\cal RQ }$,
 initially in a pure entangled  state $|\psi^{\cal RQ } \rangle$.
 The initial density operator of the system ${\cal Q }$
 is then obtained from that of ${\cal  RQ}$  by a partial trace over 
 the reference system ${\cal R}$:
 $\rho^{\cal Q}={\rm Tr}_{\cal R} [|\psi^{\cal RQ }
 \rangle  \langle \psi^{\cal RQ } | ]$.
 The system's state $\rho^{\cal Q}$ is then encoded by using 
 an ancillary system $\cal A$
 that consists of two qubits for code 1 (see Fig.~1 in Ref.~\cite{Cory98})
 and a single qubit for code 2
 (see Fig.~6 in Ref.~\cite{DarrigoBenentiFalci07_2}). 
 The system and the ancillary qubits are then transmitted in 
 $N=3$ (for c1) or $N=2$ (for c2) channel uses; 
 system ${\cal R}$ remains ideally isolated from any environment. 
 Then the receiver applies the decoding operation 
 ${\cal D}^{\cal QA}$ to system $\cal QA$. After 
 tracing out $\cal A$, we obtain the final, generally mixed 
 state of system ${\cal RQ}$:
 \begin{equation}
  \rho^{\cal RQ'}=\textrm{tr}_{\cal A}\big[\openone^{\cal R} \otimes 
                 {\cal D}^{\cal QA} 
                  \circ {\cal E}_N^{\cal QA} \big(|\tilde{\psi}^{\cal RQA}\rangle 
                  \langle\tilde{\psi}^{\cal RQA}| \big)\big],
 \end{equation}
 where $|\tilde{\psi}^{\cal RQA}\rangle$ denotes the initial state of 
 $\cal RQA$ after encoding. The entanglement fidelity $F_e$ is 
 given by the fidelity between the initial and the recovered 
 state of $\cal RQ$: 
 \begin{equation}
  F_e=\langle\psi^{\cal RQ}| \rho^{\cal RQ'} |\psi^{\cal RQ}\rangle. 
 \end{equation}

\begin{figure}
   \includegraphics[height=8.cm]{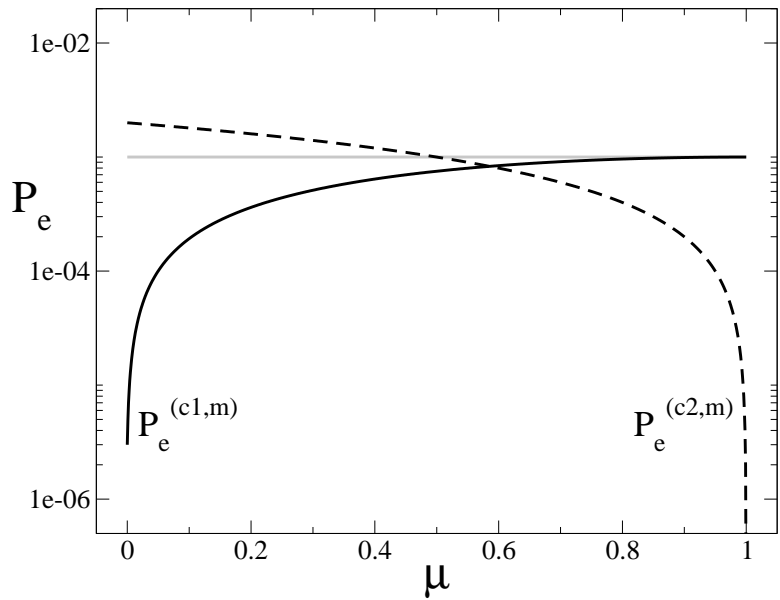}
   \caption{Plot of the transmission error probability $P_\mathrm{e}=1-F_e$ 
            as a function of the memory factor $\mu$, for the  
            transmission of a single qubit ($P_\mathrm{e}=1-p_0=10^{-3}$, 
            gray line), the three-qubit code
            ($P_\mathrm{e}^{(c1,m)}$, full curve)
            and the two-qubit code
            ($P_\mathrm{e}^{(c2,m)}$, dashed curve).}
   \label{figure}
\end{figure}

 The sharing of maximally entangled (Bell) states between two communicating 
 parties is a very interesting physical problem that can be conveniently 
 formulated using the above picture. $F_e$ just gives the 
 probability that codes c1 or c2 are successful. The three-qubit code c1
 works when the channel introduces at most a phase-flip error.
 In the memoryless case ($\mu=0$) this happens with probability  
 $F_e^{(c1)}=-2p_0^3 + 3p_0^2$ (see Refs.~\cite{NielsenChuang},\cite{BenentiCasatiStrini}). 
 Memory changes this probability, and the entanglement fidelity reads 
 \begin{equation}
  F_e^{(c1, m)}=p_0q_0^2+p_0q_0r_0+p_0r_0r_z+p_zr_zq_0
     =F_e^{(c1)}-\mu(2-\mu)(F_e^{(c1)}-p_0),
  \label{Fe_code1}
 \end{equation}
 where $r_{0,z} \equiv 1-q_{0,z} $ are the conditional  probabilities that the
 channel acts on two subsequent qubits via a different Pauli operator.
 The first term in the right hand side of (\ref{Fe_code1})
 is the probability that no 
 errors occur during the transmission, while the other terms
 correspond to a single phase error
 during the first, the second, or the third qubit transmission.
 $F_e^{(c1,m)}$ is a monotonously decreasing 
 function of $\mu$, that is, memory degrades the performance of the 
 three-qubit code (see Fig.~\ref{figure}). It is worth noticing that 
 for any $\mu$ the entanglement fidelity of the three-qubit code 
 is better than for the simple transmission of the system qubit
 (in this latter case $F_e=p_0$). It is interesting to consider the 
 case of a small error probability $\epsilon\equiv 1-p_0\ll 1$.
 Here the memoryless three-qubit code makes the transmission error 
 probability $P_\mathrm{e}=1-F_e \propto \epsilon^2$, while 
 memory restores the $\epsilon$-dependence: 
 \begin{equation}
  P_\mathrm{e}^{(c1)} \simeq 3\epsilon^2\quad \Longrightarrow \quad 
  P_\mathrm{e}^{(c1,m)} \simeq \mu(2-\mu)\epsilon.
  \label{errorprobabilityc1}
 \end{equation}
 To grasp the implications of this remark, let us assume 
 that we have a dephasing channel 
 characterized by an error probability $\epsilon=10^{-3}$. 
 At $\mu=0$ the three-qubit code 
 drastically lowers this error to 
 $P_\mathrm{e}^{(c1)} \simeq 3 \cdot 10^{-6}$. 
 However, a weak memory is sufficient to significantly degrade the performance of the code.
 For instance, $P_\mathrm{e}^{(c1,m)} \simeq 2 \cdot 10^{-4}$ at 
 $\mu=0.1$.

 Now we turn to the simple coding-decoding scheme discussed in 
 Ref.~ \cite{DarrigoBenentiFalci07_2}. This two-qubit code is designed 
 to take advantage of correlations:  memory 
 enhances the probability that
 the same operator ($\openone$ or $\sigma_z$) acts 
 on two qubits subsequently transmitted down the channel.
 Therefore, c2  encodes a qubit in the subspace spanned by $\{|01\rangle, 
 |10\rangle\}$, which is noiseless with respect to the application of 
 $\openone \otimes \openone$ or $\sigma_z \otimes \sigma_z$. The entanglement
 fidelity in presence of memory is given by
 \begin{equation}
   F_e^{(c2,m)}=p_0q_0+(1-p_0)q_z=F_e^{(c2)}+\mu(1-F_e^{(c2)}),
  \label{Fe_code2}
 \end{equation}
 where $F_e^{(c2)}$ is the entanglement fidelity of c2 at $\mu=0$.
 A first question is whether this code improves the entanglement fidelity 
 with respect to the simple transmission of the system qubit. 
 This is the case provided that  
 $\mu > (2p_0-1)/2p_0$. This condition is fulfilled for any $p_0$ 
 when $\mu>0.5$ and $F_e^{(c2,m)}\to 1$ when $\mu\to 1$
 (see again Fig. \ref{figure}). 
 For small dephasing probabilities $\epsilon$ we obtain 
 \begin{equation}
  P_\mathrm{e}^{(c2,m)} \simeq 2(1-\mu)\epsilon.
  \label{errorprobabilityc2}
 \end{equation}
 Note that $P_\mathrm{e}^{(c2,m)}<P_\mathrm{e}^{(c1,m)}$ when 
 $\mu >2-\sqrt{2} \simeq 0.6$. 

\section{Final remarks}
 In the case of small dephasing probabilities, $\epsilon\ll 1$, 
 the quantum capacity (\ref{capacitymarkov}) is weakly
 affected by memory. On the other hand, in this regime small
 values of the memory factor $\mu$ are sufficient to have a significantly 
 detrimental impact on the three-qubit code. The two-qubit code
 solves the problem only in the case of strong memory effects. 
 It would be interesting to investigate codes working on a small number
 of qubits and suitable for the regime of weak noise and memory,
 $\epsilon \ll 1$, $\mu\ll 1$.

\section*{Acknowledgments}
A.D. and G.F. acknowledge support from the EU-EuroSQIP (IST-3-015708-IP) and
MIUR-PRIN2005 (2005022977).



\begin{thebibliography}{0}

\bibitem{NielsenChuang} M. A. Nielsen and I. L. Chuang,
	 {\it Quantum Computation and Quantum Information}
	 (Cambridge Unuversity Press, Cambridge, 2000).
\bibitem{BenentiCasatiStrini} 
         G. Benenti {\it et al}.,
         {\it Principles of Quantum Computation and Information},
         Vol. I: Basic concepts (World Scientific, Singapore, 2004);
         Vol. II: Basic tools and special topics
         (World Scientific, Singapore, 2007).
\bibitem{BarnumNielsenSchumacher98} H. Barnum {\it et al}., 
	 {\it Phys. Rev. A} {\bf 57} (1998) 4153.
\bibitem{kn:solid_state_environment_noise}
         Y. Makhlin {\it et al}., {\it Rev. Mod. Phys.} {\bf 73} (2001) 357;
         E. Paladino {\it et al}., {\it Phis. Rev. Lett.} {\bf 89} (2002) 228304;
         G. Falci {\it et al}., {\it Phys. Rev. Lett.} {\bf 94} (2005) 167002;
         G. Ithier {\it et al}., {\it Phys. Rev. B} {\bf 72} (2005) 134519.
\bibitem{Banaszek04} K. Banaszek {\it et al}., 
  	 {\it Phys. Rev. Lett} {\bf 92} (2004) 257901.
\bibitem{BowenMancini04} G. Bowen and S. Mancini, 
         {\it Phys. Rev. A} {\bf 69} (2004) 012306; 
\bibitem{MemoryChannel}
         V. Giovannetti, {\it J. Phys. A: Math. Gen.} {\bf 38} (2005) 10989;
         N. Datta and T. Dorlas, {\it quant-ph/0712.0722}.
\bibitem{KretschmannWerner05} D. Kretschmann and R. F. Werner, 
         {\it Phys. Rev. A} {\bf 72} (2005) 062323.           
\bibitem{MacchiavelloPalma02} C. Macchiavello and G. M. Palma, 
	 {\it Phys. Rev. A} {\bf 65} (2002) 050301.
\bibitem{Cory98} D. G. Cory {\it et al}.,
	 {\it Phys. Rev. Lett.} {\bf 81} (1998) 2152. 
\bibitem{Leung99} D. Leung {\it et al}.,
	 {\it Phys. Rev. A} {\bf 60} (1999) 1924. 
\bibitem{QECCcorrelations} 
         R. Klesse and S. Frank, 
         {\it Phys. Rev. Lett.} {\bf 95} (1996) 230503; 
         L.-M. Duan and G.-C. Guo,  
         {\it Phys. Rev. A} {\bf 59} (1999) 4058; 
         E. Novais {\it et al}., 
         J. P. Clemens {\it et al}., 
         {\it Phys. Rev. A} {\bf 69} (2004) 062313;
         {\it Phys. Rev. Lett.} {\bf 97} (2006) 040501; 
         A. Shabani, {\it quant-ph/0703142}.
\bibitem{Hamada} H. Hamada, 
         {\it J. Math. Phys.} {\bf 43} (2002) 4382.  
\bibitem{PlenioVirmani07_1} M. B. Plenio and S. Virmani, 
         {\it Phys. Rev. Lett.} {\bf 99} (2007) 120504.
\bibitem{DarrigoBenentiFalci07_1} A. D'Arrigo {\it et al}.,
         {\it New J. Phys.} {\bf 9} (2007) 310.
\bibitem{DarrigoBenentiFalci07_2} A. D'Arrigo {\it et al}.,
        {\it quant-ph/0710.3472}, {\it Eur. Phys. J. Special Topics}
        (in press).
\bibitem{PlenioVirmani07_2} 
         M. B. Plenio and S. Virmani, {\it quant-ph/0710.3299}.
\bibitem{CodingTheorem}
         I. Devetak, {\it IEEE Trans. Inf. Theory} {\bf 51} (2005) 44;
         P. Hayden {\it et al}., {\it quant-ph/0702005};
         R. Klesse, {\it Phys. Rev. A} {\bf 75} (2007) 062315.
\end{thebibliography}
\end{document}